\documentclass[superscriptaddress,showpacs,preprintnumbers,preprint]{revtex4}
\usepackage{amsmath,amssymb,graphicx}

\begin{document}

\newcommand*{\pku}{School of Physics and State Key Laboratory
of Nuclear Physics and Technology, \\Peking University, Beijing
100871, China}\affiliation{\pku}
\newcommand*{\CHEP}{Center for High Energy
Physics, Peking University, Beijing 100871,
China}\affiliation{\CHEP}

\title{First-Digit Law in Nonextensive Statistics\footnote{Published in Phys. Rev. E82, 041110 (2010)}}

\author{Lijing Shao}\affiliation{\pku}
\author{Bo-Qiang Ma}\email[Corresponding author.
Electronic
address:~]{mabq@pku.edu.cn}\affiliation{\pku}\affiliation{\CHEP}

\begin{abstract}
Nonextensive statistics, characterized by a nonextensive parameter
$q$, is a promising and practically useful generalization of the
Boltzmann statistics to describe power-law behaviors from physical
and social observations. We here explore the unevenness of the first
digit distribution of nonextensive statistics analytically and
numerically. We find that the first-digit distribution follows
Benford's law and fluctuates slightly in a periodical manner with
respect to the logarithm of the temperature. The fluctuation
decreases when $q$ increases, and the result converges to Benford's
law exactly as $q$ approaches $2$. The relevant regularities between
nonextensive statistics and Benford's law are also presented and
discussed.
\end{abstract}

\pacs{02.50.Cw, 05.20.-y}

\maketitle

\section{Introduction}

The foundation of statistical physics was established at the end of
19th century by Maxwell, Boltzmann, and
Gibbs~\cite{m60a,m60b,b77,g02}. The celebrated
Maxwell-Boltzmann-Gibbs statistics underpins revolutionary concepts
such as ergodicity and stosszahlansatz, which means that all admissible
microstates are equiprobable over a long period of time. Cooperative
effects and nonlinear dynamics are important in leading to
sufficient statistics. The Boltzmann distribution has a
distinguished exponential behavior proportional to $e^{-\beta E}$,
where $E$ is the energy, and $\beta =1/k_BT$ is the reciprocal of
the temperature $T$ with the Boltzmann constant $k_B$.

Recently, it was found that the Boltzmann statistics follows a peculiar
digit law~\cite{sm10b}, named Benford's law, which is also called
the first-digit law or the significant digit law~\cite{n81,b38}. The
law states that the occurrence of the nonzero leftmost digit, {\it
i.e.}, $1,2,...,9$, of numbers from many real world sources is not
uniformly distributed as one might naively expect, but instead, the
nature favors smaller ones according to a logarithmic formula,
\begin{equation}\label{benford}
P_{\rm Ben}(d) = \log_{10} \left( 1+ \frac{1}{d} \right), \quad
d=1,2,...,9,
\end{equation}
where $P_{\rm Ben}(d)$ is the probability of a number having the
first digit $d$. The first-digit distribution of the Boltzmann
statistics fluctuates slightly around Benford's law in a periodical
manner with respect to the logarithm of the temperature of the
system~\cite{sm10b}. Moreover, two quantum distributions, {\it
i.e.}, the Bose-Einstein statistics and the Fermi-Dirac statistics,
are also proven to comply with Benford's law in analogous
manners~\cite{sm10b}.

On the other hand, the natural and social numbers are not always
discovered to follow the exponential Boltzmann distribution, other
than that, many sources are actually found to demonstrate power-law
behaviors. In 1988, Tsallis proposed an entropy formula with a
nonextensive parameter $q$ to describe the prevalent power-law
behaviors~\cite{t88,t99,t09a,t09b}. In the limit of $q \rightarrow
1$, it recovers the familiar Boltzmann statistics. This novel
statistics is named the Tsallis statistics, or nonextensive
statistics. It has several well-defined mathematical rigidities and
performs impressively well in various practical domains.
Empirically, the Tsallis statistics is widely employed to explain
data from nonextensive viewpoints in
various aspects in physics, chemistry, economics, computer science,
biology, cellular automata, self-organized criticality, scale-free
networks, linguistics, and other sciences (see Chap. $7$ in
Ref.~\cite{t09b} and references therein for extensively concrete
examples). Theoretically, the nonextensiveness can be explained in
terms of fluctuations of temperature~\cite{ww00}, or
superstatistics~\cite{bc03}. The physics behind nonextensiveness is
believed to be long-range correlations, strongly quantum
entanglements, and deformed phase space due to insufficient
statistics in space and/or time~\cite{t99,t09a,t09b}.

For the comprehensive existence of the Tsallis statistics, it
appears intriguing to look into its digit distribution, besides the
studied canonical ones~\cite{sm10b}. The influence from the
nonextensive parameter $q$ on results may reveal further
regularities of natural statistics. We study the first-digit
distribution of nonextensive statistics in detail both
analytically and numerically. We find that, analytically, in the
range of $1 \leq q<2$, it slightly fluctuates around the first-digit
law in a periodical manner with respect to the logarithm of the
temperature. The deviation from Benford's law is diminutive. As
$q$ varies monotonously from $1$ to $2$, the amplitude of
fluctuation becomes smaller and smaller, and in the limit of $q
\rightarrow 2$, nonextensive statistics conforms to Benford's law
exactly. Hence, it explains the underlying reason why many sources
from systems with nonextensiveness respect the first-digit law in
an almost precise way.

The paper is organized as follows. In Sec. II,
nonextensive statistics is briefly reviewed and the normalized
probability density is presented. In Sec.~\ref{result}, we make
the analytical and numerical comparisons between the first-digit
distribution of nonextensive statistics and the significant digit
law. Then, several relevant insights are discussed in
Sec.~\ref{discussion}, including scale invariance,
base invariance, and mantissa distribution. Section~\ref{sum}
summarizes the results of the paper.

\section{nonextensive statistics}

Inspired by the probabilistic description of multifractal
geometries, Tsallis postulated a possible generalization of
entropy~\cite{t88},
\begin{equation}\label{qentropy}
S_q = k_B \frac{1-\sum_{i=1}^{W} p_i^q}{q-1}, \quad q \in \mathcal
{R},
\end{equation}
where $q$ characterizes the nonextensiveness of the considered
system and $\{p_i\}$ are the probabilities associated with
$W\in\mathcal {N}$ microscopic configurations, satisfying
$\sum_{i=1}^{W} p_i =1$. In the limit of $q\rightarrow1$, $S_q$
elegantly recovers the conventional Boltzmann entropy,
$S_{q\rightarrow1} = -k_B \sum_{i=1}^W p_i \ln p_i$.

The entropy in Eq.~(\ref{qentropy}) can be rewritten with the
help of $q$ algebra~\cite{t09a,t09b},
\begin{equation}
S_q = k_B \sum_{i=1}^{W} p_i \ln_q (1/p_i) = -k_B \sum_{i=1}^{W}
p_i^q \ln_q p_i = -k_B \sum_{i=1}^{W} p_i \ln_{2-q} p_i,
\end{equation}
where the $q$-logarithmic function is defined as
\begin{equation}
\ln_q x \equiv \frac{x^{1-q}-1}{1-q}, \quad q\in \mathcal{R},
\end{equation}
and when $q$ approaches $1$, $\ln_{q\rightarrow1} x = \ln x$.

In contrast to the conventional entropy, $S_q$ is non-additive for
two independent subsystems $A$ and $B$ when $q \neq 1$,
\begin{equation}
\frac{S_q(A+B)}{k_B} = \frac{S_q(A)}{k_B} + \frac{S_q(B)}{k_B} +
(1-q) \frac{S_q(A)}{k_B} \cdot \frac{S_q(B)}{k_B}.
\end{equation}
It is clearly seen that the deviation of $q$ from $1$ reflects the
nonextensiveness of relevant systems.

Through the entropic maximizing procedure, the distribution function
$f_q(E;\beta)$ can be obtained~\cite{t88,t99,t09a,t09b},
\begin{equation}
f_q(E;\beta) \propto e_q^{-\beta E} = \left[ 1 - (1-q)\beta E
\right]^{\frac{1}{1-q}},
\end{equation}
where the $q$-exponential function is defined as
\begin{equation}
e_q^x \equiv [1+(1-q)x]^{\frac{1}{1-q}}, \quad 1+(1-q)x>0.
\end{equation}
Note that when $q$ approaches $1$, $f_q(E;\beta)$ returns to the
standard Boltzmann distribution proportional to
$e_{q\rightarrow1}^{-\beta E} = e^{-\beta E}$.

After normalization to unit, $f_q(E;\beta)$ is written as
\begin{equation}\label{f}
f_q(E;\beta) = \beta(2-q) \cdot \left[ 1 - (1-q)\beta E
\right]^{\frac{1}{1-q}}, \quad 1\leq q<2.
\end{equation}
For ranges other than $1\leq q<2$, we are not considering in this
paper for reasons listed below.
\begin{enumerate}
\renewcommand{\labelenumi}{(\roman{enumi})}
\item When $q<1$, there exists an upper limit for the energy, $E_{\rm
upper}= [(1-q)\beta]^{-1}=k_BT/(1-q)$, whose physical meaning is not
well understood yet.
\item While $q \geq 2$, $f_q(E;\beta)$
cannot be normalized, because the power $1/(1-q)$ becomes larger
than $-1$.
\end{enumerate}

\section{The first-digit distribution of Tsallis statistics}

\label{result}

With the knowledge of nonextensive statistics well-prepared from
the above section, we are going to explore its first-digit
distribution $P_q(d;\beta)$, which now depends on the nonextensive
parameter $q$, besides the temperature $T$. Utilizing the language
of probability density, the likelihood to have the first digit $d$
equals~\cite{sm10b}
\begin{equation}\label{pd}
P_q(d;\beta) = \sum_{n = -\infty}^{\infty} \int_{d \cdot
10^n}^{(d+1) \cdot 10^n} f_q(E;\beta) {\rm d} E.
\end{equation}

Combining Eqs.~(\ref{f}) and (\ref{pd}), we can get $P_q(d;\beta)$
straightforward,
\begin{equation}
P_q(d;\beta) = \sum_{n = -\infty}^{\infty} \left\{ \left[1-10^n\beta
d(1-q)\right]^{\frac{q-2}{q-1}} - \left[1-10^n\beta
(d+1)(1-q)\right]^{\frac{q-2}{q-1}}\right\}.
\end{equation}

A nice property of the Boltzmann-Gibbs statistics and the
Fermi-Dirac statistics, is still preserved~\cite{sm10b},
\begin{equation}\label{scaleinv}
P_q(d;10\beta) = P_q(d;\beta),
\end{equation}
which origins from the multiplication appearance of $E$ and $\beta$
in Eq.~(\ref{f}). As to be explained in Sec. IV,
Eq.~(\ref{scaleinv}) is closely related to the scale invariance of
Benford's law. Therefore, we can define a new function~\cite{sm10b},
\begin{equation}
P_q^\star(d;\alpha) = P_q(d;\beta=10^\alpha),
\end{equation}
which appears to be a one-periodical function, {\it i.e.},
$P_q^\star(d;\alpha) = P_q^\star(d;\alpha+1)$.

By analogy with the Boltzmann distribution~\cite{sm10b}, we expand
$P_q^\star(d;\alpha)$ into Fourier series,
\begin{equation}
P_q^\star(d;\alpha) = \sum_{n=-\infty}^{\infty} c_n
e^{i\cdot2n\pi\alpha}.
\end{equation}
We denote the constant component of the Fourier series of
$P_q^\star(d;\alpha)$, $c_0$, as a new function $P_q(d)$, and
through nontrivial calculations resembling Ref.~\cite{sm10b}, we can
get
\begin{eqnarray}
P_q(d) &\equiv& c_0 = \int_0^1 P_q^\star(d;\alpha) {\rm d} \alpha \\
\nonumber
    &=& \int_0^1 \sum_{n = -\infty}^{\infty} \left\{
    \left[1-10^{n+\alpha}
d(1-q)\right]^{\frac{q-2}{q-1}} - \left[1-10^{n+\alpha}
(d+1)(1-q)\right]^{\frac{q-2}{q-1}}\right\} {\rm d} \alpha \\
\nonumber
    &=& \int_{-\infty}^{+\infty} \left\{
    \left[1-10^{\alpha}
d(1-q)\right]^{\frac{q-2}{q-1}} - \left[1-10^{\alpha}
(d+1)(1-q)\right]^{\frac{q-2}{q-1}}\right\} {\rm d} \alpha \\
\nonumber
    &=& \int_0^\infty \left\{ \left[ 1-ud(1-q)\right]^{\frac{q-2}{q-1}}
- \left[ 1-u(d+1)(1-q)\right]^{\frac{q-2}{q-1}}\right\}
\frac{1}{u\cdot{\rm ln}10} {\rm d} u,
\end{eqnarray}
where a substitution $u=10^\alpha$ is adopted. Now, the differential
equation of $P_q(d)$ with respect to $d$ is rather concise,
\begin{equation}
P^\prime_q(d) = \frac{1}{{\rm ln}10} \left( \frac{1}{d+1} -
\frac{1}{d}\right).
\end{equation}

By utilizing the normalization condition, $\sum_{d=1}^9 P_q(d)=1$,
we attain
\begin{equation}\label{c0}
P_q(d) = \log_{10}\left(1+\frac{1}{d}\right) \equiv P_{\rm Ben}(d),
\end{equation}
which turns out to be the first-digit law exactly, in analogy with
three canonical statistics~\cite{sm10b}. Therefore, we conclude
that, at a fixed $q$, the first-digit distribution of nonextensive
statistics fluctuates around the significant digit law periodically,
with respect to $\alpha$, or equivalently, the logarithm of $\beta$.
Furthermore, we stress that, the central value, {\it i.e.}, $c_0$,
is independent of the nonextensive parameter $q$. In contrast, as
we will see later, the strength of fluctuation depends on $q$.

\begin{figure}
\includegraphics[width=10cm]{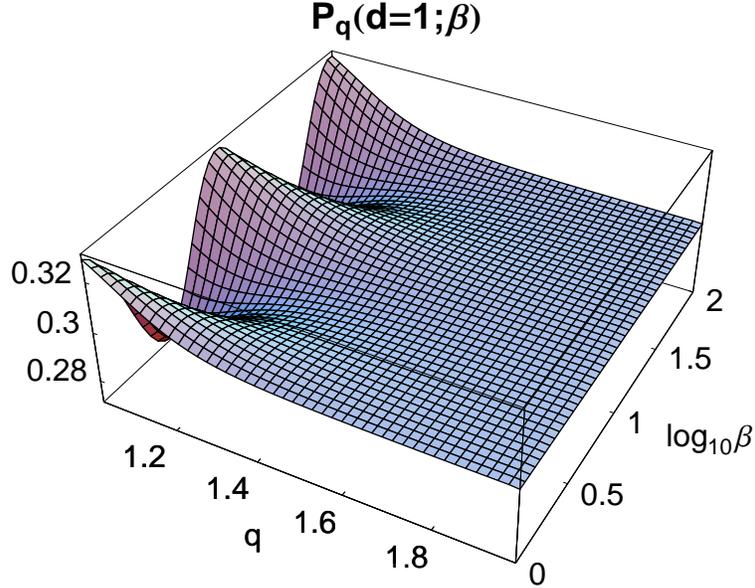}
\caption{(Color online). The probability of the Tsallis statistics
with the first digit $d=1$, {\it i.e.}, $P_q(d=1;\beta)$, versus the
logarithm of the reciprocal of the temperature $\beta = 1/k_BT$ and
the nonextensive parameter $q$.\label{d1}}
\end{figure}

Numerically, the probability of owning the first digit $d=1$, {\it
i.e.}, the function $P_q(d=1;\beta)$, is illustrated in
Fig.~\ref{d1}. The cross section of the figure at $q=1$ is the very
case for the Boltzmann statistics~\cite{sm10b}. The maximum
amplitude to deviate from Benford's law is less than $0.03$ when
$q=1$~\cite{sm10b}. For $q\neq1$, from the figure, we see that the
deviation is even smaller and decreases monotonously with increasing
$q$. In the limit of $q\rightarrow2$, we prove in Sec. IV
analytically that the digit distribution of nonextensive statistics
complies with Benford's law exactly.

\begin{figure}
\includegraphics[width=8cm]{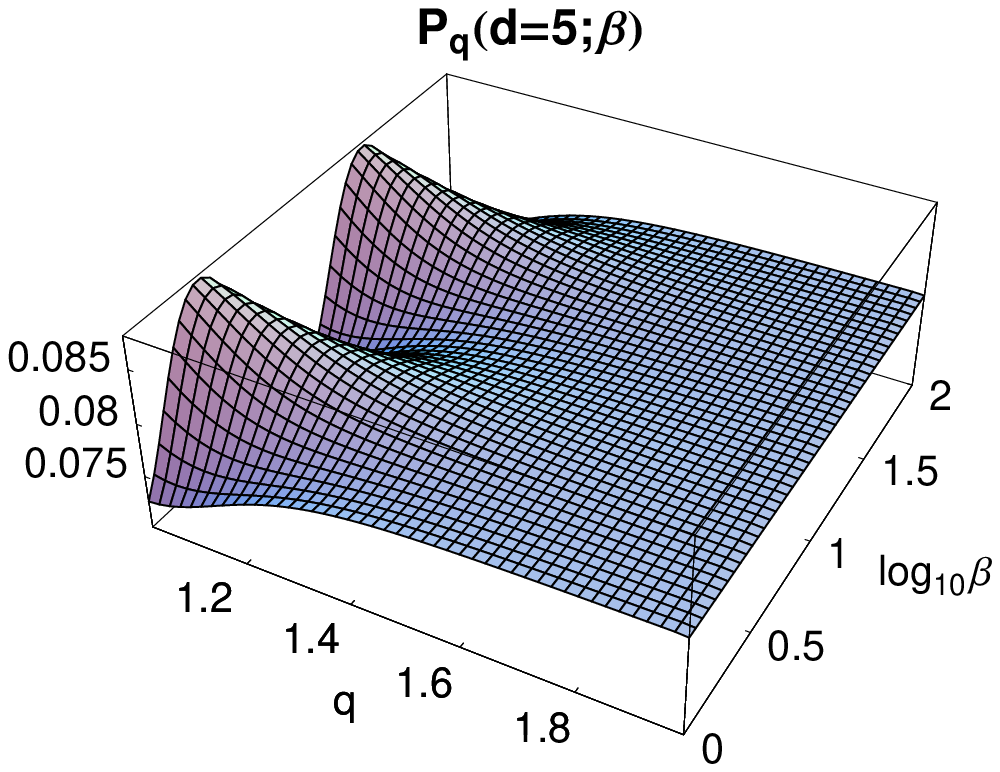}
\includegraphics[width=8cm]{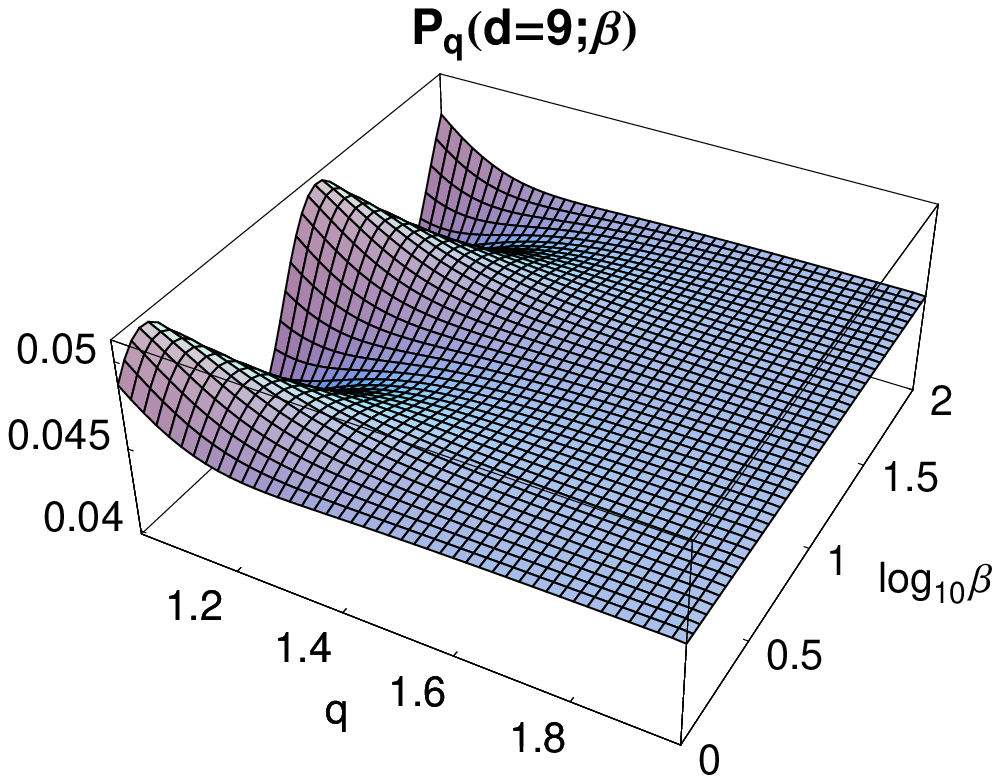}
\caption{(Color online). The probabilities of the Tsallis statistics
with first digit $d=5$ (left) and $d=9$ (right), versus the
logarithm of the reciprocal of the temperature $\beta = 1/k_BT$ and
the nonextensive parameter $q$.\label{d5d9}}
\end{figure}

The probabilities for digits other than digit $1$ behave similarly.
In Fig.~\ref{d5d9}, we depict the probabilities of possessing the
first digit as $d=5$ and $d=9$ as examples, versus the nonextensive
parameter $q$ and the logarithm of $\beta$. The central values
follow exactly Benford's law. The deviations from Benford's law for
digits $2$-$9$ are smaller than that for digit $1$. Here for $5$
and $9$, their deviations are both less than $0.01$ for
$q=1$~\cite{sm10b}, and further smaller for $q\neq1$.

\section{Discussions}

\label{discussion}

Benford's law applies to numerous data from natural sources, {\it
e.g.}, areas of lakes, lengths of rivers~\cite{b38}, physical
constants~\cite{bk91}, various quantities of pulsars~\cite{sm10a},
hadron full widths~\cite{sm09}, complex atomic spectrum~\cite{p08}, and
$\alpha$-decay half lives~\cite{bmp93,nr08}. Meanwhile,
nonextensive statistics performs successfully at describing
ubiquitous power-law behaviors~\cite{t09b}, and possesses elegantly
conceptual merits as well~\cite{t88,t99,t09a,t09b}. Hence, the
regularities between Benford's law and nonextensive statistics,
especially, the influence from the nonextensive parameter $q$,
appear as an amusing and significative issue.

Generally, $q$ is explained as a parameter reflecting microscopic
mechanism of deforming phase space or fluctuating the temperature of
the system, however unfortunately, due to the imperfect
understanding of the dynamical details of generating
nonextensiveness, it cannot be determined as a prior nowadays. A
practical way to obtain $q$ is through numerical fittings to
Eq.~(\ref{f}). Our results might point to another possible way to
access $q$ by counting the occurrence of the first digits and
investigating their deviations from Benford's law, which costs less
time than conventional fitting procedure.

Now in the following, let us discuss three properties of the first
digit law related to the studies presented in this paper.

First, Benford's law is scale invariant, which is discovered by
Pinkham in 1961~\cite{p61}. This property means that the law remains
unchanged under numerical rescalings, hence does not depend on any
particular choice of units. As mentioned, in our study, it is
closely related to Eq.~(\ref{scaleinv}), which is ascribed to the
multiplication appearance of $\beta$ and $E$. The rescaling of
$\beta$ is equivalent to inversely rescaling $E$. Therefore,
concerning the scale invariance of Benford's law, we expect that
Eq.~(\ref{scaleinv}) emerges naturally. Worthy to note that,
Benford's law is also base invariant, which means that the
regularity of the law is independent of the base
$b$~\cite{h95b,h95c}. In the binary system ($b=2$), octal system
($b=8$), or other base system, the data, as well as in the decimal
system ($b=10$), all fit the general Benford's law,
\begin{equation}
P_{\rm Ben}(d) = \log_b \left( 1+\frac{1}{d}\right), \quad
d=1,2,...,b-1.
\end{equation}

Second, the independence of $P_q(d)$ on $q$ in Eq.~(\ref{c0})
breaks out as an astonishing result at first sight. However, it was
proven by Smith by utilizing the language of digital signal
processing in his textbook that, in reference to the logarithmic
coordinates, the digit distribution of any probability distribution
function should fluctuate around Benford's law with respect to the
rescaling parameter~\cite{s07}. Therefore, the $q$-independence is
expected. As for our study, as stated above, the dependence on
$\beta$ is equivalent to the rescaling of $E$. Therefore, the
oscillation of the digit distribution around Benford's law is
consistent with Smith's analysis from the viewpoint of rescaling.

The last point we would like to present is the proof of the
coincidence of the first-digit distribution of nonextensive
statistics and the significant digit law in the $q\rightarrow2$
limit. It can be ascribed to the ``$1/m$'' behavior of mantissa
distribution~\cite{sm10b}. The mantissa $m \in [\,0.1,~1\,)$ is the
significant part of a floating-point positive number $x$, defined
uniquely as $x = m \times 10^n$, where $n$ is an integer. It was
pointed out that, if the probability density of $m$ is distributed
according to $1/m$, then Benford's law is guaranteed~\cite{sm10b}.
Actually, the ``$1/m$'' distribution of mantissa is equivalent to
the $n$-digit Benford's law~\cite{sm10b}. For nonextensive
statistics, when $q$ approaches $2$, $f_{q\rightarrow2}(E;\beta)
\propto 1/(1+\beta E)$. The integral divergence occurs when $E
\rightarrow +\infty$, where the behavior of $f_q(E;\beta)$ is
proportional to $1/E$. Thus, all mantissa contribution comes from
the $E\rightarrow+\infty$ region with probability density
proportional to $1/m$. Consequently, nonextensive statistics
follows Benford's law exactly in the limit of $q\rightarrow2$
accordingly.

\section{Summary}

\label{sum}

The Maxwell-Boltzmann-Gibbs statistics is one of the most celebrated
achievements in the history of physics, and it has many implications
and applications in various physical as well as social domains.
However, there also exist numerous examples characterized by
power-law behaviors other than the canonical exponential
distribution. The power-law behavior is elegantly and efficiently
described by nonextensive statistics with a nonextensive parameter
$q$, proposed by Tsallis in 1988.

While the Boltzmann statistics was proven to follow Benford's
law~\cite{sm10b}, which states the uneven occurrence of the first
nonzero digit, then it becomes intriguing to look into the digit
distribution of nonextensive statistics and the dependence on the
nonextensive parameter $q$. We find analytically that the first
digit distribution of nonextensive statistics has similar behaviors
as that of the Boltzmann distribution. It fluctuates slightly and
periodically around Benford's law with respect to the logarithm of
the temperature. With increasing nonextensiveness in the range of
$1\leq q<2$, the fluctuation decreases monotonously. In the limit of
$q \rightarrow 2$, nonextensive statistics follows Benford's law
exactly. Furthermore, the fluctuations of temperature in physical
systems can smooth down the oscillation to the central value, which
corresponds exactly to the first-digit law. Therefore, we reveal
that, the frequent appearance of Benford's law in the natural and
social data is theoretically expected in systems that follow
nonextensive statistics.

\section*{Acknowledgments}

This work is partially supported by National Natural Science
Foundation of China (Nos.~11005018, 10721063, 10975003, 11035003).
It is also supported by Hui-Chun Chin and Tsung-Dao Lee Chinese
Undergraduate Research Endowment (Chun-Tsung Endowment) at Peking
University, and by National Fund for Fostering Talents of Basic
Science (Nos.~J0630311, J0730316).

\bibliography{99}

\end{document}